\documentclass[aps,prd,superscriptaddress,nofootinbib,amsmath,amsfonts,preprintnumbers,groupedaddress,showpacs,10pt,english]{revtex4}
\usepackage{amsmath}
\usepackage{amssymb}
\usepackage{babel}
\usepackage{wrapfig}
\usepackage{cancel}

\usepackage{relsize,exscale}
\makeatletter

\usepackage{array,multirow,graphicx}
\usepackage{soul}
\usepackage{dcolumn}
\usepackage{newlfont}
\usepackage{bm}
\usepackage[colorlinks,citecolor=blue,urlcolor=blue,linkcolor=blue]{hyperref}
\usepackage[figtopcap]{subfigure}
\usepackage{color}
\usepackage{verbatim}

\def\be{\begin{align}}
\def\ee{\end{align}}
\def\bea{\begin{eqnarray}}
\def\eea{\end{eqnarray}}
\def\bal{\begin{align}}
\def\eal{\end{align}}

\usepackage{scalerel}
\usepackage{tikz}
\usetikzlibrary{svg.path}
\definecolor{orcidlogocol}{HTML}{A6CE39}
\tikzset{
 orcidlogo/.pic={
 \fill[orcidlogocol] svg{M256,128c0,70.7-57.3,128-128,128C57.3,256,0,198.7,0,128C0,57.3,57.3,0,128,0C198.7,0,256,57.3,256,128z};
 \fill[white] svg{M86.3,186.2H70.9V79.1h15.4v48.4V186.2z}
 svg{M108.9,79.1h41.6c39.6,0,57,28.3,57,53.6c0,27.5-21.5,53.6-56.8,53.6h-41.8V79.1z M124.3,172.4h24.5c34.9,0,42.9-26.5,42.9-39.7c0-21.5-13.7-39.7-43.7-39.7h-23.7V172.4z}
 svg{M88.7,56.8c0,5.5-4.5,10.1-10.1,10.1c-5.6,0-10.1-4.6-10.1-10.1c0-5.6,4.5-10.1,10.1-10.1C84.2,46.7,88.7,51.3,88.7,56.8z};}}
\newcommand\orcid[1]{\href{https://orcid.org/#1}{\mbox{\scalerel*{
\begin{tikzpicture}[yscale=-1,transform shape]
\pic{orcidlogo};
\end{tikzpicture}
}{|}}}}
\graphicspath{{./}{Figs/}}
\begin{document}

\date{\today}

\title{ 3-dimensional charged   black holes in $f({Q})$ 
gravity}

\author{G.~G.~L.~Nashed }\email{nashed@bue.edu.eg}

\affiliation {$^1$ Centre for Theoretical Physics, The British University, P.O. 
Box 43, El Sherouk City, Cairo 11837, Egypt }

\author{Emmanuel N. Saridakis }\email{msaridak@noa.gr}

\affiliation {National Observatory of Athens, Lofos Nymfon, 11852 Athens,
Greece}
\affiliation {
 CAS Key Laboratory for Researches in Galaxies and Cosmology,
Department of Astronomy, University of Science and Technology of China, Hefei,
Anhui 230026, P.R. China}
\affiliation {
 Departamento de Matem\'{a}ticas, Universidad Cat\'{o}lica del
Norte, Avda. Angamos 0610, Casilla 1280 Antofagasta, Chile}

\begin{abstract}
We present new exact charged black hole solutions in (2+1) dimensions  
within the framework of  $f({Q})$ gravity, where ${Q}$ denotes the non-metricity scalar.   By considering a cubic $f({Q})$ form  we derive  classes of charged and uncharged spherically symmetric solutions, and
we identify conditions under which these reduce to the well-known Banados-Teitelboim-Zanelli (BTZ) black hole. Notably, our analysis reveals a novel charged solution that is asymptotically Anti-de Sitter (AdS) but cannot be continuously deformed into a GR counterpart, and thus arising purely from the
 higher-order nature of the non-metricity corrections. Furthermore, we explore the geometrical  properties of the solutions, demonstrating the existence of multiple horizons and a softer central singularity compared to GR.  Additionally, we calculate various thermodynamic quantities, such as Hawking temperature, entropy, and heat capacity, with results confirming   thermal stability. Finally, a detailed study of the geodesic motion and the effective potentials unveils stable photon orbits, as well as   the effect of cubic non-metricity corrections on orbital dynamics.

\end{abstract}

\maketitle

\section{Introduction}\label{S1}

Multiple cosmological observations~\cite{SupernovaSearchTeam:1998fmf,Planck:2015mvg,Planck:2015bpv,
Addazi:2021xuf,CosmoVerse:2025txj} suggest the need to modify General Relativity (GR) at fundamental geometric levels \cite{CANTATA:2021asi}. GR, based on Riemannian geometry, uses the Levi-Civita connection and identifies the Ricci curvature scalar $R$ as the Lagrangian. However, this formulation excludes both torsion and non-metricity from its geometric structure. An alternative description of gravity arises in teleparallel gravity, where torsion $T$ replaces curvature as the gravitational Lagrangian \cite{Aldrovandi:2013wha,Cai:2015emx}.  More recently, a third geometrical formulation, known as symmetric teleparallel gravity, was developed by Nester~\cite{Nester:1998mp}, where gravity is described through non-metricity rather than curvature or torsion.

Symmetric teleparallel gravity has attracted growing interest as a foundation for modified gravity theories. In this framework, the gravitational dynamics is embedded in the non-metricity scalar
${Q}$, and the connection is flat and torsion-free. This approach introduces new geometrical degrees of freedom and opens new ways for   cosmological models and gravitational solutions \cite{BeltranJimenez:2017tkd, BeltranJimenez:2020lee}.

One of the most promising extensions of this theory is
$f({Q})$ gravity \cite{Harko:2018gxr}, in which the Lagrangian is promoted from a linear to an arbitrary function of ${Q}$. This generalization enables richer phenomenology and has been applied to a wide range of contexts, including cosmology and astrophysics \cite{Harko:2018gxr,Barros:2020bgg}, particularly in the modeling of compact stars  \cite{Frusciante:2021sio, Mandal:2020buf}. Despite this progress, most investigations remain focused on four-dimensional spacetimes, while the lower-dimensional applications, especially in (2+1) dimensions, remain comparatively underexplored. Such reduced-dimensional models are not only mathematically tractable but also yield exact solutions that provide information on  quantum aspects of gravity and strong-field matter configurations  \cite{Carlip:1994tt,Carlip:1994gy,Carlip:1994hq, Banados:1992wn}.

Extensive research has investigated $f({Q})$ gravity in numerous scenarios, including constraints from observational data~\cite{Lazkoz:2019sjl,Ayuso:2020dcu,Anagnostopoulos:2021ydo}, cosmography \cite{Mandal:2020buf} energy conditions ~\cite{Mandal:2020lyq}, bouncing scenarios ~\cite{Mandal:2021wer,Bajardi:2020fxh}, black holes~\cite{DAmbrosio:2021zpm}, and the growth index of matter perturbations~\cite {Nguyen:2021cnb}. Comprehensive reviews and theoretical developments in this area can be found in Refs.~\cite{Esposito:2021ect,Atayde:2021pgb,Kimura:2020mpk,Dimakis:2021gby}.  Significantly,
 $f({Q})$ gravity  can account for the accelerated expansion of the universe, with observational precision comparable to other well-established modified gravity theories, and as it was shown recently, it is in high agreement with the new DESI DR2 datasets \cite{Yang:2024kdo,Yang:2025mws}. This leads to degeneracy among models, revealing the importance of identifying distinct astrophysical or gravitational   signatures that can break this degeneracy \cite{Mann:2018xkm,LIGOScientific:2016aoc}. Recent contributions  have focused on such issues in various contexts  \cite{Errehymy:2022gws,Maurya:2022cyv,Tangphati:2021tcy}.  In summary, within $f({Q})$ gravity  one can obtain   interesting phenomenology \cite{BeltranJimenez:2019tme,Anagnostopoulos:2021ydo,Lazkoz:2019sjl,Lu:2019hra,Mandal:2020buf,Khosravi:2021csn,Frusciante:2021sio,Ferreira:2022jcd,Gadbail:2022jco,Sarmah:2023oum,Khyllep:2021pcu,Barros:2020bgg,
De:2022jvo,Solanki:2022ccf,DAmbrosio:2021zpm,Li:2021mdp,Dimakis:2021gby,Kar:2021juu,Wang:2021zaz,Quiros:2021eju,Mandal:2021bpd,Albuquerque:2022eac,Anagnostopoulos:2022gej,Arora:2022mlo,Pati:2022dwl,Dimakis:2022wkj,DAgostino:2022tdk,Narawade:2022cgb,Emtsova:2022uij,Bahamonde:2022cmz,Sokoliuk:2023ccw,Shaikh:2023tii,Dimakis:2023uib,Koussour:2023rly,Najera:2023wcw,Atayde:2023aoj,Shabani:2023xfn,Tayde:2023xbm,De:2023xua,Junior:2024xmm,Millano:2024rog,Yang:2024tkw,Guzman:2024cwa,Alwan:2024lng,Shaily:2024tmx,Wang:2024eai,Moreira:2024unj,Dubey:2024gxa,Mushtaq:2025rfn,Nashed:2025usa,Papagiannopoulos:2025uix,Moreira:2025rxp,Dimakis:2025jrl,Carloni:2024ybx,Basilakos:2025olm} (for a comprehensive study see \cite{Heisenberg:2023lru}).

In this work  we derive   exact charged and uncharged spherically symmetric solutions
 in a (2+1)-dimensional spacetime within the context of $f({Q})$ gravity.  
 While most studies focus on standard 4-dimensional solutions \cite{DAmbrosio:2021zpm,Gogoi:2023kjt, Javed:2023vmb, Junior:2023qaq, Sokoliuk:2023pby, Wu:2024vcr, Javed:2024trj,  Al-Badawi:2024iqv, Vacaru:2025kpj}, extracting 3-dimensional solutions is both interesting and necessary, for the reasons described above. Additionally, although many studies use   linear or quadratic forms $f({Q})$, we will consider a cubic modification, since it introduces richer nonlinear dynamics and allows for novel gravitational and electromagnetic interactions. Moreover, the presence of an electric field is relevant especially  in lower dimensions, where charge-geometry couplings often yield exact and physically insightful black hole or stellar-type configurations.  We will first extract the solutions, and then we will examine their geometrical and physical characteristics in detail, including asymptotic structure, singularity behavior, and thermodynamic properties. These solutions may be useful for further investigations in quantum gravity, black hole thermodynamics, and holographic dualities within non-Riemannian geometric settings.

The structure of the present study is organized as follows. In Section \ref{Sec:II}, we introduce the basics of the mathematical formalism used in the charged version of
$f({Q})$ gravity. This section also includes the derivation of the field equations for a spherically symmetric (2+1)-dimensional spacetime, leading to the relevant differential equations. We solve these equations under two specific conditions. The first corresponds to the uncharged case, where we demonstrate that the general solution within the cubic form of $f({Q})$ reproduces the Banados-Teitelboim-Zanelli (BTZ)  black hole of GR, with the cosmological constant emerging as a function of the cubic model parameters. The second case addresses the charged scenario, where we obtain a novel black hole solution characterized by a non-trivial profile of the non-metricity scalar
${Q}$.

In Section \ref{S4} we investigate the properties of the solutions. 
In particular, in subsection \ref{S4A} 
we analyze 
the behavior of the metric functions associated with the charged solution, and we show that the spacetime is asymptotically Anti-de Sitter (AdS). We also evaluate curvature and non-metricity invariants, comparing them to those in GR, and we reveal that our solution exhibits a milder singularity than the charged  BTZ black hole \cite{Banados:1992wn}. Subsection \ref{S41} is devoted to the thermodynamic properties of the charged black hole, and we calculate the Hawking temperature, entropy, and heat capacity, demonstrating that the solution represents a thermodynamically stable configuration. 
In Subsection \ref{sec:44}  we examine the geodesic structure of the black hole, we derive the corresponding effective potential, and we confirm the existence of stable photon orbits. Subsection \ref{multi} explores the possibility of constructing black-hole solutions with multiple horizons based on our charged model. Finally, the concluding Section \ref{Conclusions} summarizes the main findings and discusses the physical implications of our results.

\section{Mathematical Formalism}\label{Sec:II}
 
\subsection{Covariant Formulation of $f({Q})$ Theory}
\label{Sec:IIA}

In a general geometric framework, the affine connection can be decomposed into three components: the Levi-Civita connection $\left(\left\{_{~~\mu \nu}^\lambda\right\}\right)$, the contortion connection $\left(K_{~~\mu \nu}^\lambda\right)$ and the disformation tensor $\left( L_{~~\mu v}^\lambda\right)$ \cite{Heisenberg:2023lru}. This decomposition is given by:
    \begin{equation}\label{eq1}
    \Gamma_{~~\mu \nu}^\lambda= \left\{_{~~\mu \nu}^\lambda \right\}+K_{~~\mu \nu}^\lambda+L_{~~\mu \nu}^\lambda,
    \end{equation}
where  the components are defined as follows:
\begin{eqnarray}
   && \left\{_{~~\mu \nu}^\lambda \right\} = \frac{1}{2} g^{\lambda \alpha}\left(\partial_\mu g_{\alpha \nu}+\partial_\nu g_{\alpha \mu}-\partial_\alpha   g_{\mu \nu}\right),\quad
    K_{~~\mu \nu}^\lambda = \frac{1}{2}\left(\mathbb{T}_{~~\mu\nu}^\lambda+\mathbb{T}_{\mu~~\nu}^{~~\lambda}+\mathbb{T}_{\nu~~\mu}^{~~\lambda}\right),\quad L_{\ \mu \nu }^{\lambda}= \frac{1}{2}({Q}_{~~\mu\nu }^{\lambda}-{Q}_{\mu~~\nu }^{\ \lambda}-{Q}_{\nu~~\mu }^{\ \lambda}).\nonumber\\
    &&
\end{eqnarray}
Here, $\mathbb{T}^{\lambda}_{~~\mu\nu}$ denotes the torsion tensor, while ${Q}_{\lambda\mu\nu}$ represents the nonmetricity tensor. Both are constructed using the metric tensor $g_{\mu\nu}$ and the affine connection $\Gamma^{\lambda}_{~~\mu\nu}$ as:
\begin{eqnarray}\label{eq2}
    \mathbb{T}^{\lambda}_{~~\mu\nu} := \Gamma^{\lambda}_{~~\mu\nu}-\Gamma^{\lambda}_{~~\nu\mu},\quad\quad
    {Q}_{\lambda\mu\nu} := \nabla_{\lambda}g_{\mu\nu} = \partial_{\lambda}g_{\mu\nu}-\Gamma^{\alpha}_{~~\lambda\mu}g_{\alpha\nu}-\Gamma^{\alpha}_{~~\lambda\nu}g_{\alpha\mu}.
\end{eqnarray}
The nonmetricity scalar ${Q}$ is defined as ${Q} = Q_{\lambda \mu \nu} P^{\lambda\mu\nu}$, where $P^{\lambda}{~\mu\nu}$ is its conjugate, expressed by:
\begin{equation}\label{eq3}
   P^{\lambda}_{~~\mu\nu} = -\frac{1}{4}{Q}^{\lambda}_{~\mu \nu} + \frac{1}{4}\left({Q}^{~\lambda}_{\mu~\nu} + {Q}^{~\lambda}_{\nu~~\mu}\right) + \frac{1}{4}{Q}^{\lambda}g_{\mu \nu}- \frac{1}{8}\left(2 \tilde{{Q}}^{\lambda}g_{\mu \nu} + {\delta^{\lambda}_{\mu}{Q}_{\nu} + \delta^{\lambda}_{\nu}{Q}_{\mu}} \right).
\end{equation}
It is worth noting that in some works there exists an alternative sign convention, namely ${Q} = -{Q}_{\lambda\mu\nu}P^{\lambda\mu\nu}$, which affects the sign of the scalar ${Q}$, a  detail that needs to be taken into account when comparing results between different studies. When ${Q}$ replaces the Ricci scalar ${R}$ in the Einstein-Hilbert action, one obtains the symmetric teleparallel equivalent of general relativity (STEGR). However, similar to GR, STEGR  cannot produce a dynamical dark-energy sector, prompting the extension to $f({Q})$ gravity, akin to the $f({R})$ modification.

In terms of coordinate transformations, the connection components can also be written as:
\begin{equation}\label{eq4}
\Gamma^{\lambda}\,_{\mu \beta }=\frac{\partial y^{\lambda}}{\partial \xi^{\rho }}\partial _{\mu }\partial _{\beta }\xi ^{\rho },
\end{equation}
where $\xi^{\lambda}(y^{\mu})$ denotes an invertible function, and $\frac{\partial y^{\lambda}}{\partial \xi^{\rho}}$ is the inverse Jacobian matrix. This formulation enables a coincident gauge in which the connection components vanish, simplifying the covariant derivative to a partial derivative: $Q_{\lambda \mu \nu} = \partial_\lambda g_{\mu \nu}$. As a result, the Levi-Civita connection is related to the disformation   as $\left\{^{\lambda}_{~~\mu \nu}\right\} = -L^{\lambda}_{~~\mu \nu}$.

Now, in order to incorporate the electromagnetic field we consider the 3-dimensional action \cite{BeltranJimenez:2017tkd,Nashed:2024ush,Nashed:2024jmw}:
\begin{equation}
\label{eq5}
    S = \int \frac{1}{2\kappa^2}f({Q})\sqrt{-g}~d^{3}x + \int \mathcal{L}_{\text{em}}\sqrt{-g}~d^{3}x,
\end{equation}
with $\kappa^2$ the gravitational constant.
The electromagnetic Lagrangian is defined as $\mathcal{L}_{\text{em}} = -\frac{1}{2}F \wedge ^{\star}F$, where $F = d\varphi$ is the field strength 2-form derived from the gauge potential 1-form $\varphi = \varphi_\mu dx^\mu$. Varying the action with respect to the metric yields the field equations 
\begin{eqnarray}\label{eq6}
&&  0=  \frac{2}{\sqrt{-g}}\nabla_{\lambda}\left(\sqrt{-g}f_{{Q}}P^{\lambda}_{~~\mu\nu}\right) - \frac{1}{2}g_{\mu \nu}f + f_{{Q}}(P_{\mu\lambda\alpha}{Q}^{~~\lambda \alpha}_{\nu} - 2{Q}_{\lambda \alpha \mu}P^{\lambda \alpha}_{~~~\nu}) +\frac{1}{2}\kappa^2{{{\mathfrak{
T}}^{{}^{{}^{^{}{\!\!\!\!\scriptstyle{em}}}}}}}^\nu_\mu,\\
&&0=\partial_\nu \left( \sqrt{-g} F^{\mu \nu} \right)=0,
\end{eqnarray}
with  the energy-momentum tensor of the electromagnetic field   given by 
\begin{equation}\label{12a}
{{{\mathfrak{
T}}^{{}^{{}^{^{}{\!\!\!\!\scriptstyle{em}}}}}}}_{\iota\vartheta} = \frac{1}{4\pi} \bigg( F^{\nu}_\iota
F_{\vartheta\nu} - \frac{1}{3} g_{\iota\vartheta} F_{\mu\nu}
F^{\mu\nu} \bigg).
\end{equation}
In four dimensions this formulation has been the focus of  various studies, including those on geodesic deviation and cosmological behavior. The covariant form of the field equations reads as
\begin{eqnarray}\label{eq7}
&&0 = f_{{Q}} \mathring{G}{\mu\nu} + \frac{1}{2} g{\mu\nu}({Q} f_{{Q}} - f) + 2 f_{{Q}{Q}} P^{\lambda}{~~\mu\nu} \mathring{\nabla}\lambda {Q} + \frac{1}{2}\kappa^2 \mathfrak{T}^{\text{em}}{\mu \nu},\\
&& 0 = \partial\nu (\sqrt{-g} F^{\mu \nu}),
\end{eqnarray}
where $f_{{Q}} = \frac{df}{d{Q}}$ and $\mathring{G}_{\mu\nu}$ is the standard Einstein tensor based on the Levi-Civita connection. When $f({Q})$ is linear in ${Q}$, the theory reduces to GR.
Finally, varying the action  with respect to the connection leads to the   equation  
\begin{equation}\label{eq8}
    \nabla_{\mu}\nabla_{\nu}\left(\sqrt{-g}f_{{Q}}P^{\mu\nu}_{~~~\lambda}\right)=0.
\end{equation}

\subsection{Spherically symmetric spacetime}

The spherically symmetric  metric in (2+1) dimensions takes the form
\begin{equation}\label{1}
ds^{2}=-\mu(r)dt^{2}+\frac{dr^{2}}{\nu(r)}+r^{2}d\phi^{2}.
\end{equation}
Inserting it in the general field equations of the previous subsection provides
 the
Einstein-Maxwell field equations    as follows:
\begin{eqnarray}\label{fe}
&& 0=\frac {2 f_{{Q} {Q}}\mu\nu {Q}'  +f_{Q} \nu' \mu  +r \left[\mu   \left( f-{Q} f_{Q} \right)  -2\, \varphi'^{2}\nu \right] }{2r\mu },\label{2} \nonumber\\
&&0=\frac{\mu  fr-\mu  f_{{Q}}{Q}  r+f_{{Q}}\mu'   \nu -2\varphi'^{2}\nu r}{2r\mu  }\,\nonumber\\
&&0=\frac {2\,f \mu^{2}-2\,f_{Q}\,{Q}   \mu^{2 }+f_{Q}\nu'  \mu'   \mu  +2f_{Q}\,\nu   \mu''\mu  -f_{Q}\nu   \mu'^{2}+4\,\varphi'^{ 2}\mu  \nu  +2f_{{Q}{Q}} \nu \mu' {Q}' \mu  }{4  \mu^{2}}\,.
\end{eqnarray}
Furthermore, the non-metricity scalar  is given as 
\begin{equation}\label{5}
{Q}=-\frac{\nu \mu'}{r\mu},
\end{equation}
where primes represent  derivatives with respect to $r$.

In this work we consider the cubic
form of $f({Q})$ theory, namely
\begin{equation}\label{6}
f({Q})=-2\Lambda-{Q}+\frac{1}2 \alpha{Q}^2+\frac{1}3 \beta{Q}^3,
\end{equation}
where $\alpha$ and $\beta$ are dimensional constants  that have   units $[L]^{-2}$, $[L]^{-4}$, respectively, while  $\Lambda$ is the cosmological constant.  In this case $f_{{Q}}=-1+ \alpha{Q}+\beta{Q}^2$ and
$f_{{Q}{Q}}=\alpha+2\beta{Q}$. In order to recover GR from this modified
theory,  we should set  $\alpha=\beta=0$  \cite{BeltranJimenez:2019tme,Khyllep:2021pcu}. On the other hand, when $\beta=0$  we recover the quadratic form. Under the cubic assumption, and by substituting Eqs. (\ref{5}) and (\ref{6}) into (\ref{fe}), we obtain
\begin{eqnarray}\label{7}
&&0=\frac { 1}{ 2\mu^{3 }{r}^{4}} \left[12\left\{ 2r\mu \beta\, \mu'  \nu^{3}-{r}^{2} \mu^{2} \alpha\, \nu^{2} \right\} \mu'' -20\,r\beta \mu'^{3} \nu^{3}+3\left\{10 \nu' r\beta -8\beta\,\nu +3{r}^{2} \alpha \right\} \mu  \nu^{2} \mu'^{2} -6 \mu^{2}\nu \left\{3 \nu'  r  -2\nu\right\} \alpha\,r\mu'\right.\nonumber\\
 &&\left.-6\, \nu' {r}^ {3} \mu^{3}-12\,{r}^{4} \mu^{2} \left\{ \mu \Lambda+ q'^{2}\nu  \right\}\right],\\\label{8}
 &&0=\frac {10\,\beta\, \mu'^{3} \nu^{3}-6\, \mu' \nu  \mu^{ 2}{r}^{2}-9\,\alpha\, \mu'^{2} \nu^{2}\mu r-12\,{\Lambda}\, \mu^{3}{r}^{3}-12\, q'^{2}{r}^{3} \mu^{2}\nu }{12 \mu^{3}{r}^{3}},\\
\label{9}
 &&0=\frac{1}{12 \mu^{4}{r}^{3}} \left[ 6\,r\mu  \nu   \left\{3\,\beta\, \mu'^{2} \nu^{2}-\mu^{2}{r}^{2}-2\,\alpha\, \mu'r\mu  \nu  \right\} \mu''  -15\, \mu'^{4}r\beta\, \nu^{3}+9\,\mu   \left\{ {r}^{2}\alpha -{ \frac {8}{9}}\,\beta\,\nu +\frac{5}3\, \nu' r\beta \right\}\nu^{2} \mu'^{3}\right.\nonumber\\
 &&\left.+3\,r\nu   \mu^{2} \left\{ \alpha\,\nu  +{ r}^{2}-3\,r \nu' \alpha \right\}  \mu'^{2}-3\, {r}^{3} \mu^{3}\mu\nu'  - 12\,{r}^{3} \mu^{3} \left\{\mu  \Lambda- q'^{2}\nu   \right\}  \right],\\
\label{99} &&0=\frac {-2\,r\nu \varphi'' \mu  - \left[ -\mu'\nu  r+\mu   \left(  \nu' r+2\,\nu   \right)  \right] \varphi' }{2\mu^{2}r},
\end{eqnarray}
where the last equation,  Eq.~\eqref{99}, is the Maxwell field equation.
In order to proceed to the solution of the system of differential equations (\ref{7})--(\ref{99}) we will study the      uncharged and charged cases separately.

\subsection{The  uncharged case}

  In the uncharged case, namely for $q=0$, the  differential equations  (\ref{7})--(\ref{99})  accept  three solutions, two of which are imaginary, while  the real one takes the form:
  \begin{eqnarray}\label{11}
  &&\mu=\nu=\frac {{r}^{2}\Upsilon^{2/3}+20\,{r}^{2 }\beta+9\,{r}^{2}{\alpha}^{2}+3\,{r}^{2}\alpha\,\Upsilon^{1/3}+20c_1\beta\,\Upsilon^{1/3}}{20 \beta\,\Upsilon^{1/3}},
    \end{eqnarray}
     where $\Upsilon$ is defined as
       \begin{eqnarray}
   &&\Upsilon= \left( 90\alpha\beta+600\Lambda{ \beta}^{2}+27{\alpha}^{3}+10\sqrt {3600{\Lambda}^{2}{\beta }^{2}+ \left( 1080\Lambda\alpha-80 \right) \beta-27{\alpha }^{2}+324\Lambda{\alpha}^{3}}\beta \right)\,.
  \end{eqnarray}
To simplify the above solution we introduce ${3600{\Lambda}^{2}{\beta }^{2}+ \left( 1080\Lambda\alpha-80 \right) \beta-27{\alpha }^{2}+324\Lambda{\alpha}^{3}}=0$, and solving for $\beta$ we acquire
\begin{equation}\label{22}
\beta={\frac {2-27\,\Lambda\,\alpha\pm2\,\sqrt {1243\,
{\alpha}^{2}{\Lambda}^{2}-27\,\Lambda\,\alpha-729\,{\Lambda}^{3}{\alpha}^{3}}}{180{\Lambda}^{2}}}\,.
\end{equation}
Substituting  (\ref{22}) into the differential equations (\ref{7})--(\ref{99}) we find  $\Lambda$ as $\alpha=\frac{1}{9\Lambda}$, and  inserting into (\ref{22}) we finally obtain
\begin{equation}\label{33}
\alpha=\frac{1}{9\Lambda}, \qquad \qquad \beta=-\frac{1}{180\Lambda^2}\,.
\end{equation}
Hence, inserting (\ref{33}) into  (\ref{7})--(\ref{99}) we find  $\mu$ and $\nu$ as
\begin{equation}\label{sol}
\mu=\nu=-3\Lambda r^2+c_1\,,
\end{equation}
which is just the known   Banados-Teitelboim-Zanelli (BTZ) of GR \cite{Banados:1992wn}. Thus in the case of static spherically symmetric peacetime and through the use of the cubic form of $f({Q})$ we can see that the general solution of the resulting differential equations is the standard  BTZ black hole.

\subsection{The  charged case}

 In the charged case, namely for $q\neq 0$, the solution of the differential equations  (\ref{7})--(\ref{99}) becomes
  \begin{eqnarray}\label{11a}
 && \mu=\nu=-3\Lambda r^2+c_1\,, \qquad \varphi=const., 
   \end{eqnarray}
 which coincides with  the uncharged case and thus will not be further considered, and 
  \begin{eqnarray}\label{11}
  &&\mu= \frac{1}3{\frac {{r}^{2}}{{c_2}^{3}}}+{\frac {15}{8}}\,{\frac {c_2\,\sqrt [5]{2}}{{r}^{2/3}}}-\frac{5{2}^{2/5}}3\ln  \left( \frac{r}{r_0}\right) +{\frac {9}{8}}\,{\frac {{c_2}^{2}}{{r}^{4/3}}}-m, 
  \nonumber\\
  &&\nu=\mu\nu_1,\nonumber\\
&&  \nu_1=\frac{-9\,{c_2}^{3}{\Lambda}}{ \left( 1+1/2\,{\frac {c_2\,{2
}^{4/5}}{{r}^{2/3}}}+\frac{3}4{\frac {{c_2}^{2}{2}^{3/5}}{{r}^{4/3}
}} \right)^2}, 
\nonumber\\
&&\varphi=c_4-\frac{2}3\,\sqrt [5]{2}\ln \left( \frac{r}{r_0}\right)+{\frac {c_2}
{{r}^{2/3}}}+\frac{3}8\,{\frac {{c_2}^{2}{2}^{4/5}}{{r}^{4/3}}},
  \end{eqnarray}
  where $r_0$ is a length scale. In order to obtain  an asymptotically AdS  solution we must rewrite $\mu(r)$ as
\begin{eqnarray}\label{df8}
  &&\mu= -3\Lambda_{eff} r^2+{\frac {15}{8}}\,{\frac {c_2\,\sqrt [5]{2}}{{r}^{2/3}}}-\frac{5{2}^{2/5}}3\ln  \left( \frac{r}{r_0}\right) +{\frac {9}{8}}\,{\frac {{c_2}^{2}}{{r}^{4/3}}}-m, \nonumber\\
  &&\nu_1=\frac{1}{ \left( 1+\frac{1}2{\frac {c_2\,{2
}^{4/5}}{{r}^{2/3}}}+\frac{3}4{\frac {{c_2}^{2}{2}^{3/5}}{{r}^{4/3}
}} \right)^2}
, 
 \nonumber\\
  &&q=c_4-\frac{2}3\,\sqrt [5]{2}\ln \left( \frac{r}{r_0}\right)+{\frac {c_2}
{{r}^{2/3}}}+\frac{3}8\,{\frac {{c_2}^{2}{2}^{4/5}}{{r}^{4/3}}},
\end{eqnarray}
where $\Lambda_{eff}=-\frac{1}9{\frac {1}{{c_2}^{3}}}$. Hence, we deduce that in the case $c_2\neq 0$,        solution (\ref{df8}) is a new one, with no GR limit.

In summary,  as we can see solution (\ref{df8}) is  a new exact (2+1)-dimensional charged  spherically symmetric solution in   $f({Q})$ gravity under the cubic ansatz.  As we mentioned above,  this solution cannot reduce to the charged  spherically symmetric (2+1)-dimensional solution of GR since it is not valid   for $c_2=0$ (since   the cosmological constant diverges when $\alpha$ and $\beta$  are vanishing).  Therefore, solution  (\ref{df8}) is a new branch of black-hole solutions in the higher-order   non-metricity theory, with no GR analogue.


\section{Properties of the 3-dimensional black-hole  solution}\label{S4}

In this section we analyze in detail the properties of the  3-dimensional black-hole  solution obtained in the previous section.

\subsection{Metric, singularity and horizons}
\label{S4A}

\begin{figure}
\centering
\subfigure[ ]{\label{fig:R}\includegraphics[scale=0.27]{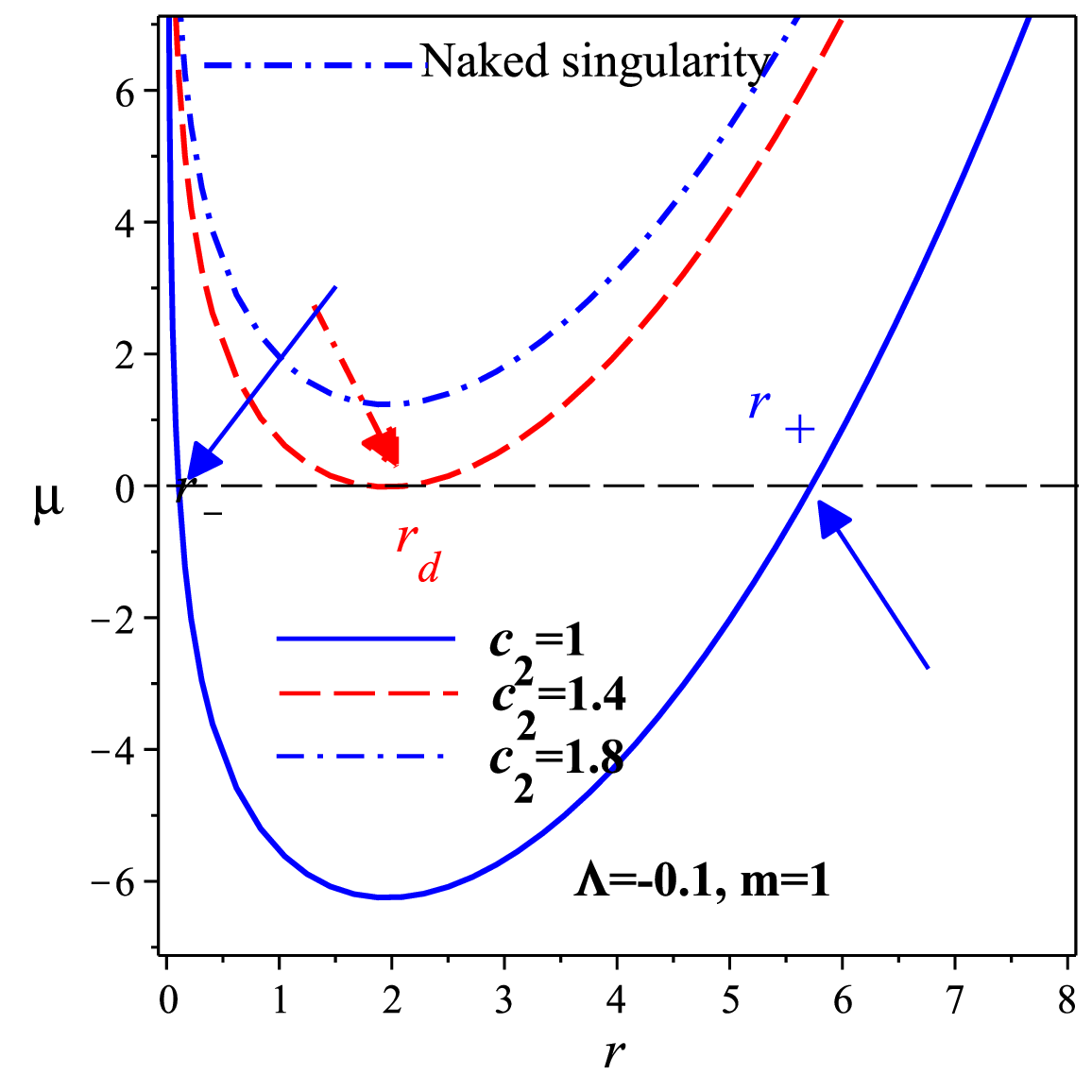}}
\hspace{1cm}  \subfigure[ ]{\label{fig:fr}\includegraphics[scale=0.27]{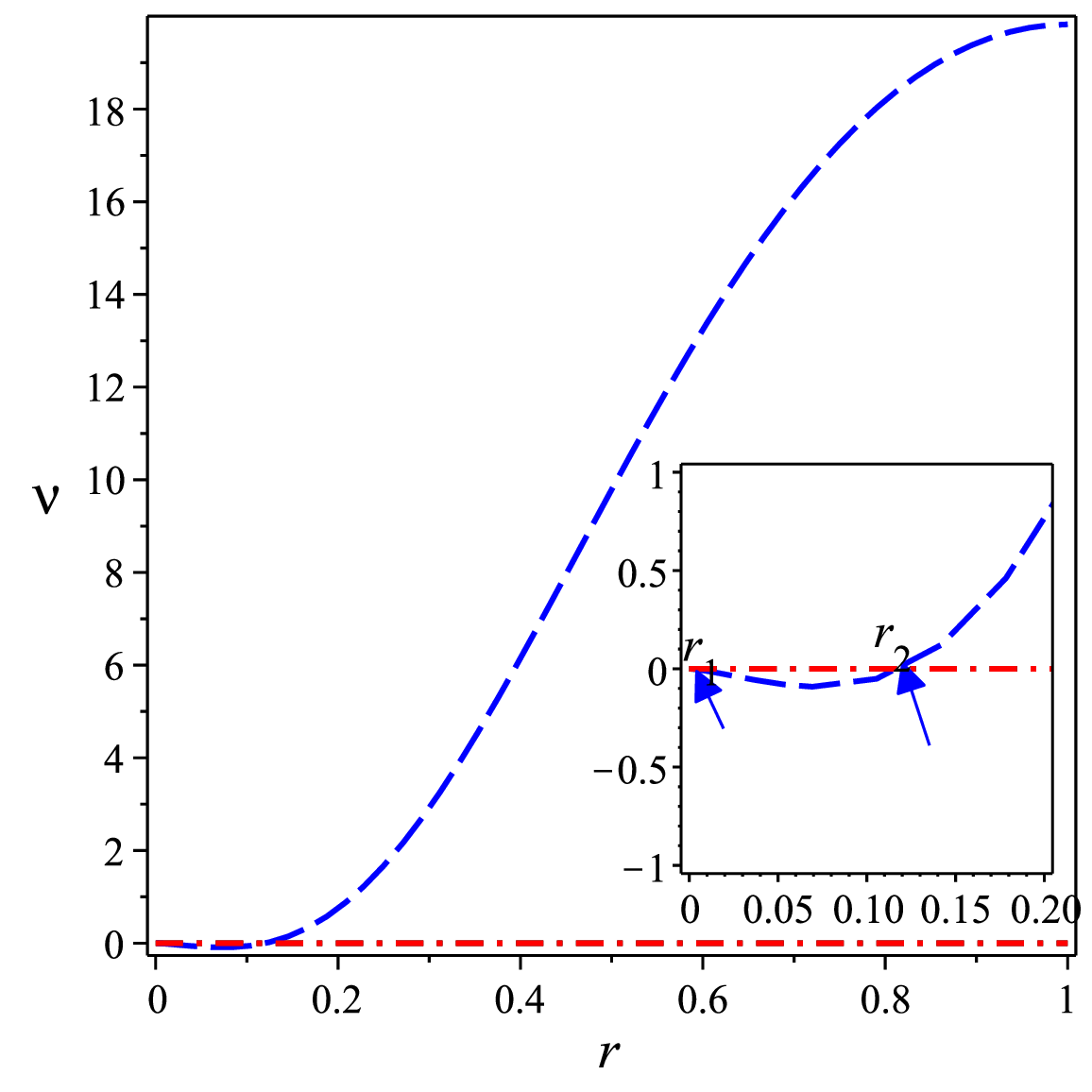}}
%
\caption[figtopcap]{\small{  \subref{fig:R} The temporal component of the metric given by Eq. (\ref{metric}).   \subref{fig:fr} The spatial component of the metric given by Eq. (\ref{metric}).}}
\label{Fig:1}
\end{figure}

The metric corresponding to the solution  (\ref{1}), after using  (\ref{df8}), takes the form
\bea\label{metric}
  && 
  \!\!\!\!\!\!\!\!\!\!\!\!\!\!\!\!\!
  ds{}^2=\Biggl[r^2 \Lambda_{eff}+{\frac {15}{8}}\,{\frac {\varphi\,\sqrt [5]{2}}{{r}^{2/3}}}-\frac{5{2}^{2/5}}3\ln  \left( \frac{r}{r_0}\right) +{\frac {9}{8}}\,{\frac {\varphi^{2}}{{r}^{4/3}}}-m\Biggr]dt^2\nonumber\\
& & -\Biggl[\frac{ \left( 1+\frac{1}2{\frac {\varphi\,{2
}^{4/5}}{{r}^{2/3}}}+\frac{3}4{\frac {\varphi^{2}{2}^{3/5}}{{r}^{4/3}
}} \right)^2}{(r^2 \Lambda_{eff}+{\frac {15}{8}}\,{\frac {\varphi \sqrt [5]{2}}{{r}^{2/3}}}-\frac{5{2}^{2/5}}3\ln  \left( \frac{r}{r_0}\right) +{\frac {9}{8}}\,{\frac {\varphi^{2}}{{r}^{4/3}}}-m)}\Biggr] dr^2+r^2d\phi^2,
\eea
where $c_2=\varphi$ and $\Lambda_{eff}=-\frac{1}{9c_2^3}$.
The behavior of   metric (\ref{metric}) is shown in Fig. \ref{Fig:1}. As we observe,   depending on the appropriate choice of the value of $\varphi$, the black hole can have 
two horizons, the inner one $r_-$ and the outer $r_+$, it can have one horizon  $r_d$ (when the two horizon   coincide), or it can exhibit a naked singularity.

Equation (\ref{metric}) shows clearly that the metric of the charged solution is asymptotically Anti-de-Sitter (AdS). Notice that there is no corresponding GR solution upon taking the limit $\beta \rightarrow 0$, namely  this charged solution has no analogue in GR. By taking the limit $\varphi \rightarrow 0$, we obtain the AdS uncharged black hole given by (\ref{sol}). Moreover, we mention  that although the metric (\ref{metric}) behaves asymptotically  as AdS charged solution, which has different $g_{tt}$ and $g^{rr}$ components, the two solutions have coinciding Killing  and event horizons.

 We proceed to the derivation of  physical singularities by calculating curvature and non-metricity scalars. In particular, we have to consider  the behavior of curvature invariants close to the roots of the function $\nu_1(r)$.  Doing so, we finally acquire
\begin{eqnarray} &&R^{\mu \nu \lambda \rho}R_{\mu \nu \lambda \rho}=R^{\mu \nu}R_{\mu \nu}=R= {Q}^{\mu \nu \lambda}{Q}_{\mu \nu \lambda}={{Q}}_{\lambda}=\tilde{{Q}}_{\lambda}={{Q}}\approx 108{\Lambda}^{2}-{\frac {168{\Lambda}^{2}\varphi{2}^\frac{4}{5}}{{r}^\frac{2}{3}}}, \end{eqnarray}
where $R^{\mu \nu \lambda \rho}R_{\mu \nu \lambda \rho}$, $R^{\mu \nu}R_{\mu \nu }$, $R$, ${{Q}}^{\mu \nu \lambda}{{Q}}_{\mu \nu \lambda}$, $Q^{\mu }Q_{\mu }$, $\tilde{{Q}}^{\mu }\tilde{{Q}}_{\mu }$  and $Q$ are the Kretschmann scalar,  the Ricci tensor square, the Ricci scalar, the non-metricity tensor square,  the non-metricity vector square and the non-metricity scalar, respectively.
 The above invariants show that:\vspace{0.1cm}\\
 a) There is a singularity at $r=0$ which is a curvature singularity. \vspace{0.1cm}\\
 b) In the charged case,  the non-metricity scalar has the form
  \begin{equation} {Q}=12\frac{ \left( 8\,{r}^{10/3}-15\,{\varphi}^{4}\sqrt [5]{2}{r}^
{2/3}-20\,{2}^{2/5}{\varphi}^{3}{r}^{4/3}-18\,{\varphi}^{5}
 \right) {\Lambda_{eff}}}{{r}^{2/3} \left( 4\,{r}^{4/3}+2\,\varphi\,{
2}^{4/5}{r}^{2/3}+3\,{\varphi}^{2}{2}^{3/5} \right) ^{2}},\end{equation} which shows that the non-metricity scalar has a singularity at $r=0$, too. Close to $r=0$,  the behavior of the Kretschmann scalar, the Ricci tensor square and the Ricci scalar for the charged solution  is given by $K=R_{\mu \nu}R^{\mu \nu}=R={Q}^{\alpha \beta \gamma}{Q}_{\alpha \beta \gamma}={Q}^{\alpha}{Q}_{\alpha}={Q}\sim r^{-2/3}$, in contrast to the  solutions of the  Einstein-Maxwell theory in both GR and non-metricity which have $K=R_{\mu \nu}R^{\mu \nu}=R={Q}^{\alpha \beta \gamma}{Q}_{\alpha \beta \gamma}={Q}^{\alpha}{Q}_{\alpha}={Q}\sim  r^{-2}$. This shows clearly that the singularity is much milder than the one obtained in GR and symmetric teleparallel gravity for the charged case.

\subsection{Thermodynamics}\label{S41}
 
In this subsection we first obtain the thermodynamical quantities
of the black-hole solution, such as   temperature and entropy, and
then we examine the validity of the first law of thermodynamics.
Applying  the Bekenstein-Hawking     relation of the black-hole temperature  
 on the event horizon $r_+$ (outermost one),
we can calculate  the Hawking temperature     through the definition of surface gravity ($\kappa$) as:
\begin{equation}
T(r_+) =\frac{\kappa}{2\pi} =\left. \frac{\mu'(r)}{4\pi}\right\vert _{r=r_{+}}=-\frac{1}{48}{\frac {72\,{\Lambda}\,{r_+}^{10/3}+15\,\varphi\,\sqrt [5]
{2}{r_+}^{2/3}+20\,{2}^{2/5}{r_+}^{4/3}+18\,{\varphi}^{2}}{\pi \,{r_+}^{7
/3}}}. \label{kGR}
\end{equation}%
 
\begin{figure}
\centering
\subfigure[~]{\label{fig:temp1}\includegraphics[scale=0.21]{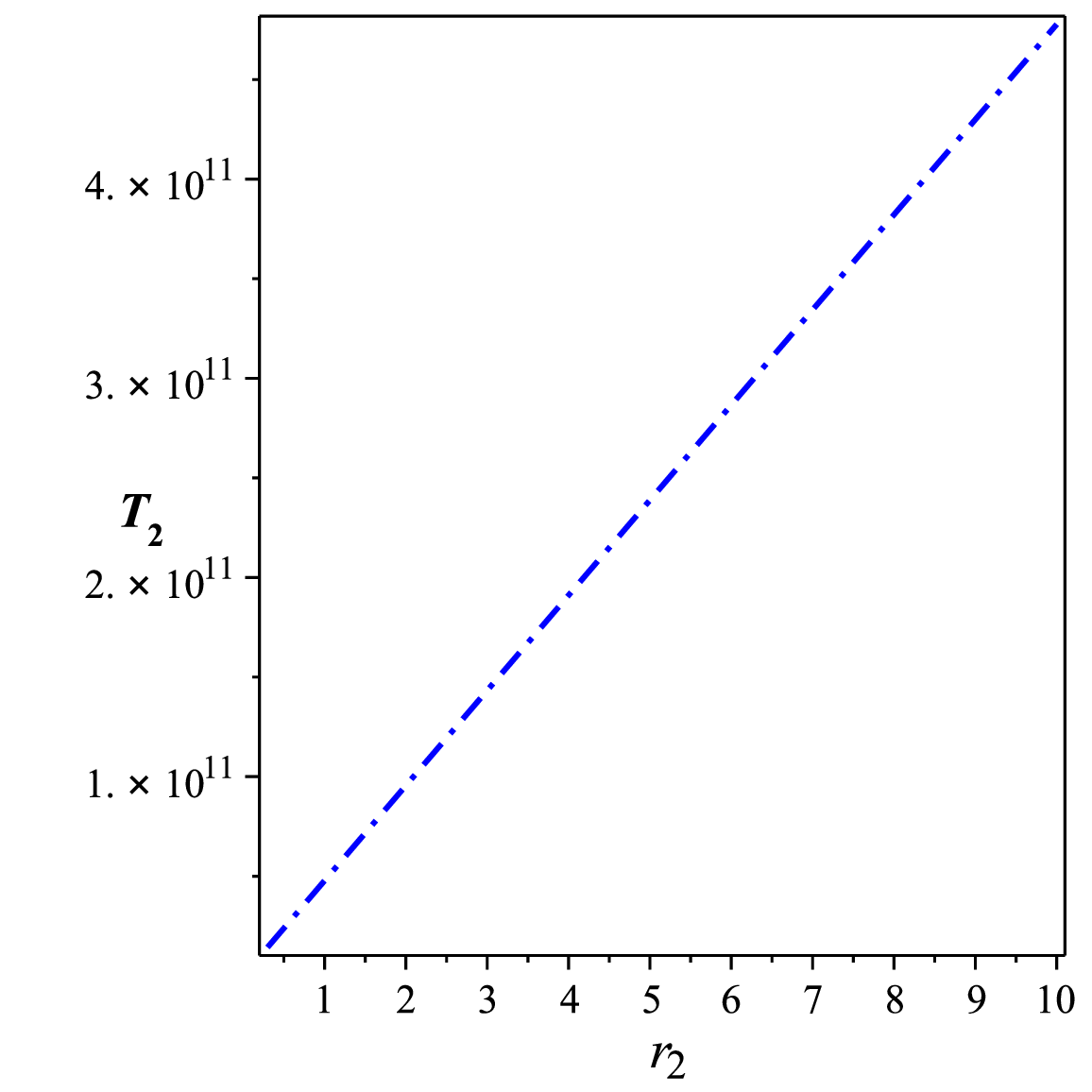}}
\subfigure[~]{\label{fig:ent1}\includegraphics[scale=0.21]{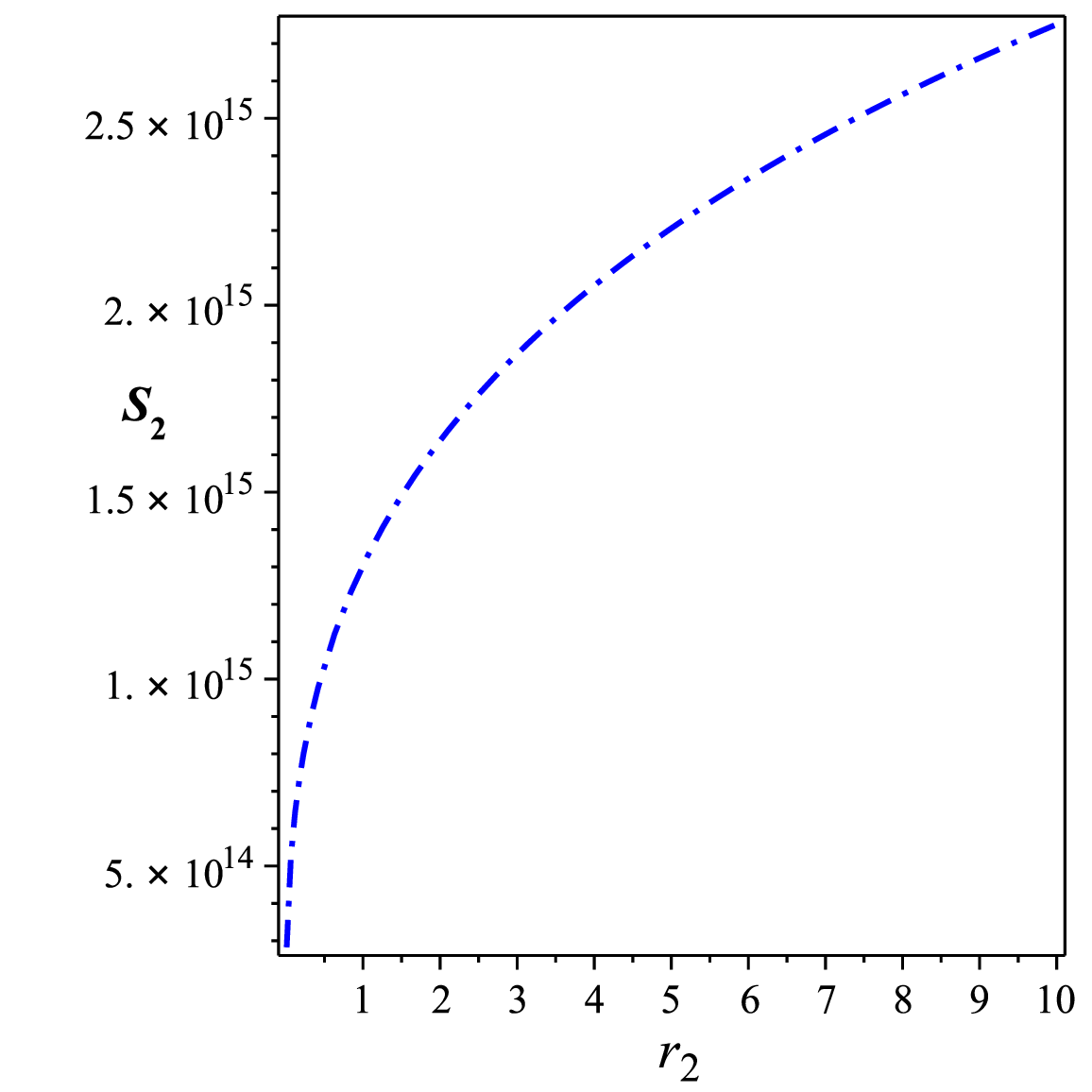}}
\subfigure[~]{\label{fig:heat1}\includegraphics[scale=0.21]{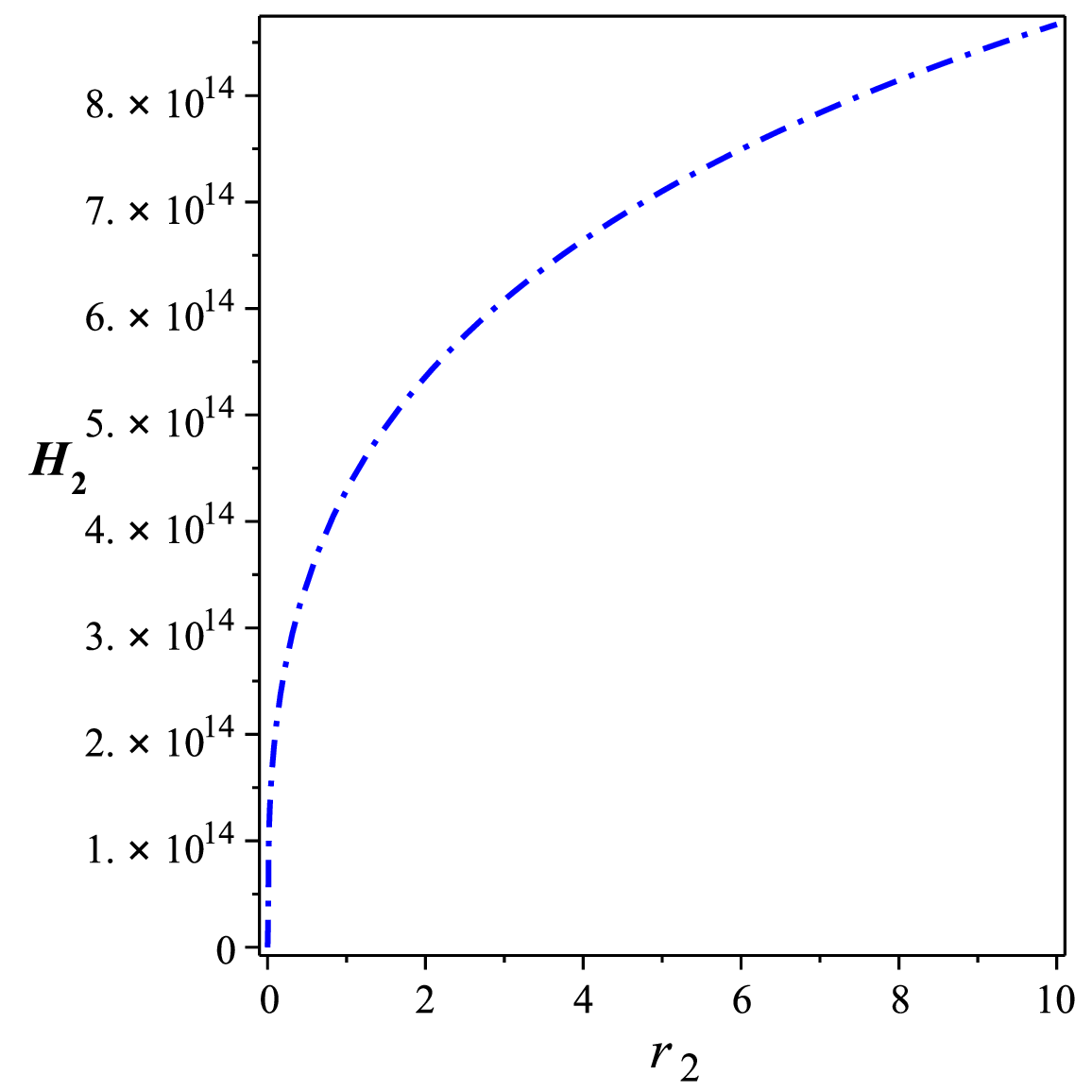}}
\subfigure[~]{\label{fig:crit}\includegraphics[scale=0.21]{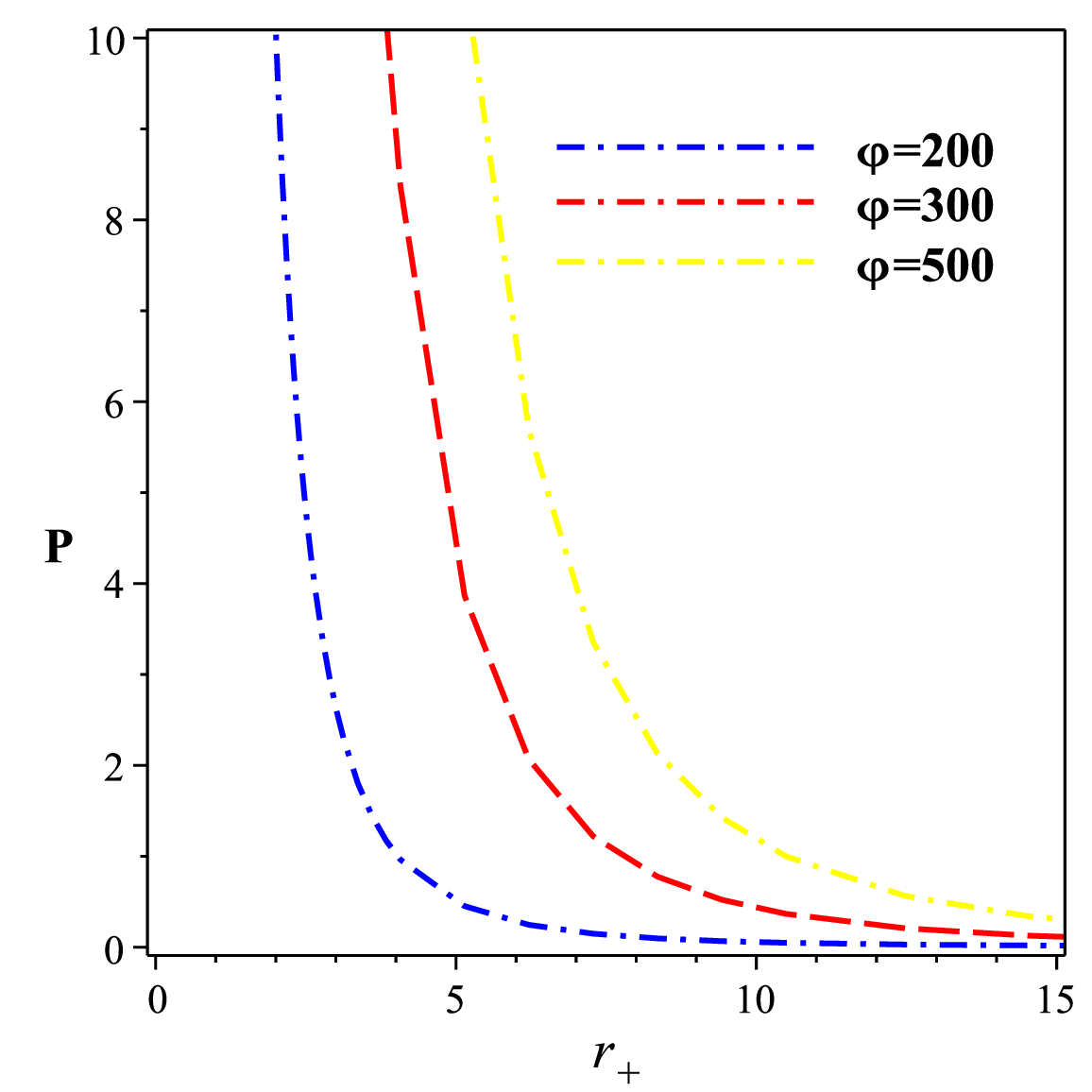}}
\caption[figtopcap]{\small{  \subref{fig:temp1} The     temperature given by Eq. (\ref{kGR}).\subref{fig:ent1} The     entropy given by Eq. (\ref{ent1}).   \subref{fig:heat1} The   heat capacity given by Eq. (\ref{heat1}). \subref{fig:crit} The       pressure given by Eq. (\ref{pre}). In all graphs we have used $\Lambda=-10^{11}$, $m=1$, $c_2=10^2$, $r_0=0.1$.  }}
\label{Fig:2}
\end{figure}
In Fig.~\ref{Fig:2}\subref{fig:temp1} we depict the behavior of the Hawking temperature given by   (\ref{kGR}),  showing that the temperature is always positive.

 In order to determine the entropy of the BTZ black hole, the area law can be applied as $S=\frac{A}{4}$  where $A$ is the horizon area     defined by $A=\left. \int_{0}^{2\pi }\sqrt{g_{\varphi \varphi }}d\varphi \right\vert
_{r=r_{+}}=\left. 2\pi \right\vert _{r=r_{+}}=2\pi r_{+}$,  where $g_{\varphi \varphi }=r^{2}$.
Hence, the semi-classical Bekenstein-Hawking entropy of the horizons is  given by 
\begin{align}\label{ent1}
&S(r_+)=\frac{{\cal A}}{4 } f_{Q}(r_+)=\frac{1}2\pi  r_+f_{Q}(r_+) \approx 4\,\pi\,r_+ \left(6\,{\Lambda}  -6\,{\frac {\varphi\,{\Lambda}\,{2}^{4/5}}{{r_+}
^{2/3}}}\right)\,,
\end{align}
with ${\cal A}=2\pi r_+$    the area of the event horizon.  The behavior  of the entropy is depicted in Fig.~\ref{Fig:2}\subref{fig:ent1},  showing that the BH solution   (\ref{ent1}) has always positive entropy.

Finally,   the heat capacity   is calculated  as \cite{Zheng:2018fyn,Nashed:2019yto,Kim:2012cma}
\begin{align} \label{heat1}
&H(r_+)=T(r_+)\left(\frac{S'(r_+)}{T'(r_+)}\right)=8\,{\frac { \left( 72\,{\Lambda}\,{r_+}^{10/3}+15\,\varphi\,
\sqrt [5]{2}{r_+}^{2/3}+20\,{2}^{2/5}{r_+}^{4/3}+18\,{\varphi}^{2}
 \right) \pi\,{\Lambda}\, \left( \varphi\,{2}^{4/5}-3\,{r_+}^{2/3
} \right) \sqrt [3]{r_+}}{-72\,{\Lambda}\,{r_+}^{10/3}+25\,\varphi
\,\sqrt [5]{2}{r_+}^{2/3}+20\,{2}^{2/5}{r_+}^{4/3}+42\,{\varphi}^{2}}}\,,
\end{align}
where $S'(r_+)$ and $T'(r_+)$ are the derivatives of entropy and Hawking temperature with respect to the outer horizon, respectively.  We draw the behavior of the heat capacity    in Fig.~\ref{Fig:2}\subref{fig:crit}, and as we observe  it has a positive pattern, ensuring that the solution under consideration is a stable one.

 We close this subsection by  investigating the critical behavior of our solution through plotting and analyzing the P-V diagrams. Here the cosmological constant $\Lambda$
 is regarded as a thermodynamic pressure $P=-\frac{\Lambda}{8\pi}$. The corresponding conjugate quantity for the pressure is the thermodynamic volume $V$, which can be obtained from the following relation:
 \begin{align}
 V=\left(\frac{\partial H}{\partial P}\right)_{S,Q}=\left(\frac{\partial M}{\partial P}\right)_{S,Q}\,,
 \end{align}
where  $M$ is the total mass and $H$ is the enthalpy of the black hole.

The critical points in the P-V diagram can be extracted  as the inflection points  from the following relations:
 \begin{align}\label{const}
 \left(\frac{\partial P}{\partial v}\right)_{v=v_c,T=T_c}=\left(\frac{\partial^ P}{\partial v^2}\right)_{v=v_c,T=T_c}=0\,,
 \end{align}
where $v$ is the specific volume.  Since we desire to compare it with the van der Waals equation, we perform  a series expansion of the latter  with the inverse of specific volume $v$, namely
\begin{equation}\label{Van}
    P=\frac T{v-b}-\frac a{v^2}\approx\frac T v+\frac{bT}{v^2}-\frac a{v^2}+O(v^{-3}).
\end{equation}
Thus, one can identify the specific volume $v$  with the horizon radius of the black holes as~\cite{Gunasekaran:2012dq,Altamirano:2013ane,Kubiznak:2012wp}
\begin{align}
v=\frac{4r_+}{d-2}\,,
\end{align}
for $d=3$.
Inserting the temperature expression (\ref{kGR})  into  (\ref{Van})  we find
\begin{align}\label{pre}
P=-{\frac {1}{576}}\,{\frac {48\,T\pi \,{r_+}^{7/3}+15\,\varphi\,\sqrt
[5]{2}{r_+}^{2/3}+20\,{2}^{2/5}{r_+}^{4/3}+18\,{\varphi}^{2}}{\pi \,{r_+}
^{10/3}}}.  
\end{align}
This equation   does not admit any inflection points and thus it does not exhibit critical behavior. This feature is obvious in Fig.~\ref{Fig:2}\subref{fig:crit}, and thus we conclude that the diagrams correspond to the ``ideal gas'' one-phase behavior. In particular, this will be clear if we try to solve Eq.~\eqref{const} to find the critical temperature and critical horizon, since it yields no real solution.
 
\subsection{Geodesics equations}\label{sec:44}

In this subsection we examine the geodesic motion of test particles around this BTZ space-time given by the metric (\ref{metric}), and we compare the results with those obtained in GR.

The geodesics equations are given by
\begin{equation}
    \ddot{x}^{\lambda}+\Gamma^{\lambda}_{\mu\nu}\,\dot{x}^{\mu}\,\dot{x}^{\nu}=0,\label{A1}
\end{equation}
where a dot represents derivative with respect to the affine parameter $\tau$.
The metric tensor $g_{\mu\nu}$ given by  (\ref{metric}) depends only on the coordinate $r$, being  independent of the coordinates $t$ and $ \phi$. Therefore, there exist  two Killing vectors $\partial_{t}=\frac{\partial}{\partial t}$ and $\partial_{\phi}=\frac{\partial}{\partial \phi}$, and the corresponding constants of motion with respect to the parameter $\tau$ can be derived using the relation $k=g_{\mu\nu}\,\xi^{\mu}\,\frac{dx^{\nu}}{d\tau}$. These constants $(\mathrm{E},\mathrm{L})$ are given by
\begin{eqnarray}
    -\mathrm{E}=g_{tt}\,\dot{t}, \qquad \qquad
    \mathrm{L}=g_{\phi\phi}\,\dot{\phi},\label{A2}
\end{eqnarray} 
where $g_{tt}$ and $g_{\phi\phi}$ are defined in  \eqref{metric}.
It is worth mentioning that since $g_{\phi\phi}$ vanishes at the BH horizon $r=r_+$ then  $\dot{t}$  diverges, while for naked singularities both  $\dot{t}$ and  $\dot{\phi}$ are positive definite as it is clear from Fig. \ref{Fig:1}. Consequently, the geodesics around a naked singularity are drastically different from those in the BH case.

The Lagrangian of the system is defined by
\begin{equation}
    \mathbb{L}=\frac{1}{2}\,g_{\mu\nu}\,\frac{dx^{\mu}}{d\tau}\,\frac{dx^{\nu}}{d\tau}.\label{A5}
\end{equation}
Using the metric   (\ref{metric})  we find
\begin{eqnarray}
    g_{tt}\,\dot{t}^2+g_{rr}\,\dot{r}^2+g_{\phi\phi}\,\dot{\phi}^2=\epsilon,\label{A6}
\end{eqnarray}
where $\epsilon=0$ for null geodesics and $\epsilon=-1$ for time-like geodesics.
Substituting the metric   components, and after some algebra, we obtain
\begin{eqnarray}
    \dot{r}^2=\frac{\epsilon}{g_{rr}}-\frac{\mathrm{E}^2}{g_{tt}g_{rr}}-\frac{L^2}{r^2g_{rr}}.\label{A7}
\end{eqnarray}
To simplify the calculations  we assume that $g_{tt}g_{rr}=-1$, which arises from the fact that $\nu_1(r)=1$ that is satisfied when the constant $\varphi=0$. When $g_{tt}g_{rr}\approx -1$  Eq. \eqref{A7} can be rewritten as:
\begin{eqnarray}
    \dot{r}^2=\frac{\epsilon}{g_{rr}}+\mathrm{E}^2-\frac{L^2}{r^2g_{rr}},\label{A77}
\end{eqnarray}
and therefore we deduce that  the effective potential of the system is given by
\begin{eqnarray}
    V_{eff} (r)&=&-\frac{\epsilon}{g_{rr}}+\frac{L^2}{r^2g_{rr}}, \label{A8}
\end{eqnarray}
where $\Lambda$ is connected with $\alpha$ and $\beta$ through $\Lambda=-\frac{1}{\sqrt{9\alpha\beta}}$. As we can see, the potential of the system is affected by the dimensional constants $\alpha$ and $\beta$ involved in the cosmological constant. Hence, this dimensional constant alters the effective potential of the system compared to the result of GR.
\begin{figure}
\centering
\subfigure[~]{\label{fig:VL}\includegraphics[scale=0.27]{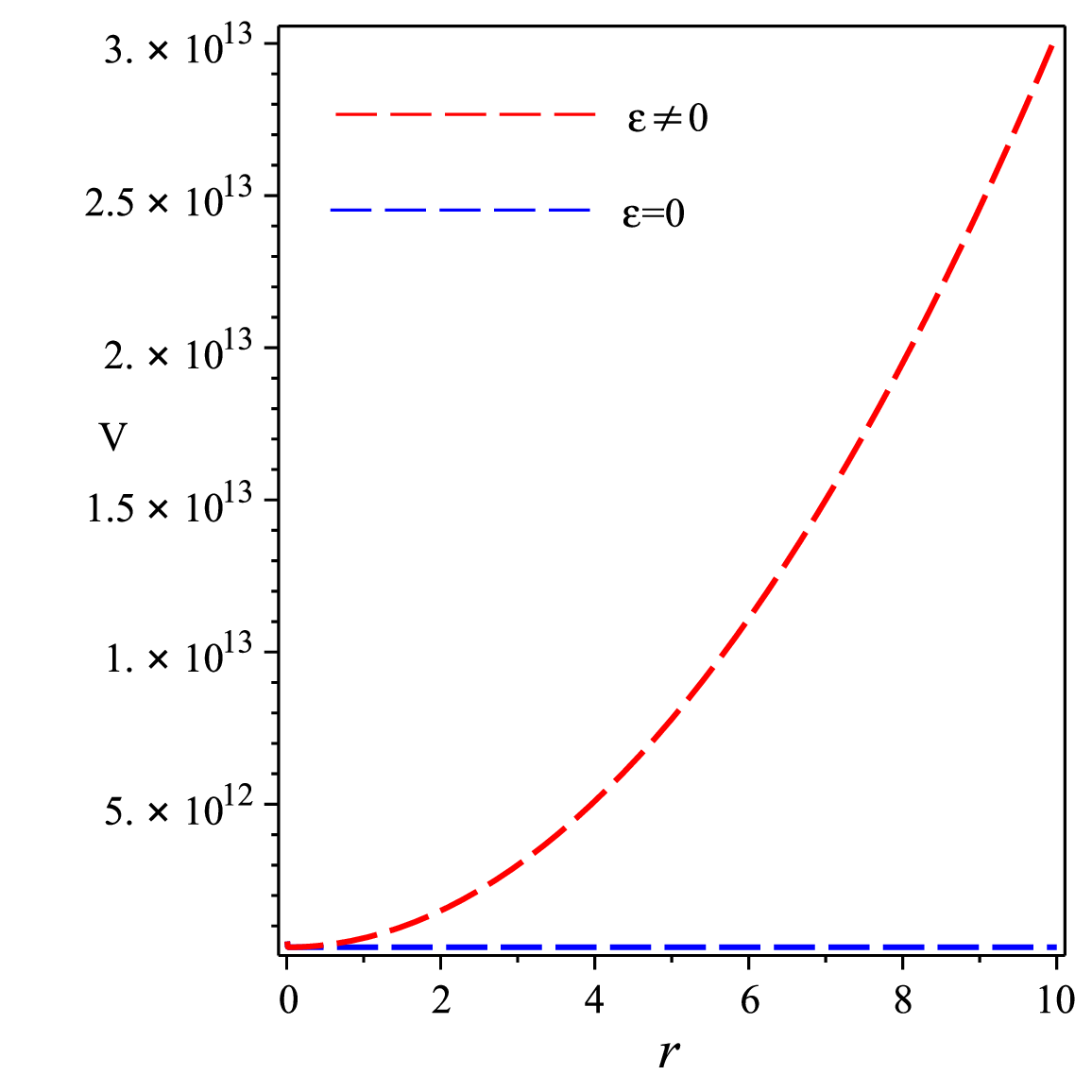}}
\hspace{1cm}
\subfigure[~]{\label{fig:ddv}\includegraphics[scale=0.27]{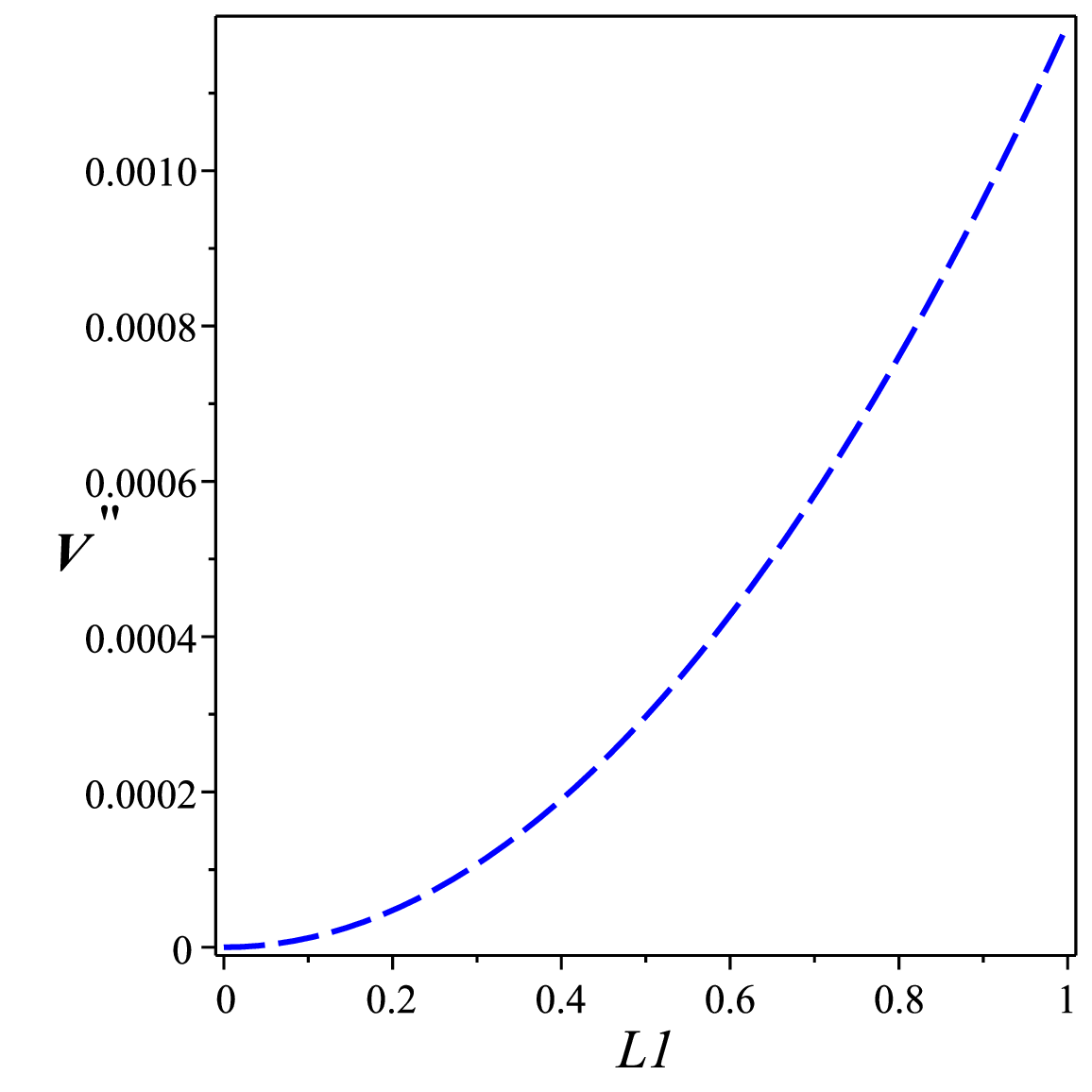}}
%
\caption[figtopcap]{\small{  \subref{fig:VL} The    potential   \eqref{A8} with $L=1$ and $\epsilon=0$,$\epsilon\neq0$.   \subref{fig:ddv} The  second derivative of the  potential     \eqref{A8} with $L=1$  and $\epsilon\neq0$.}}
\label{Fig:3}
\end{figure}

 Fig. \ref{Fig:3}  illustrates the behavior of the effective potential. In panel \subref{fig:VL} we depict the potential for  $\epsilon=0$ and $\epsilon=1$, while in panel \subref{fig:ddv}  we depict the  second derivative of the effective potential for null geodesics,  which shows the stability of the photon orbits.

\subsection{Multi horizons}\label{multi}

\begin{figure}[ht]
\subfigure[~]{\label{fig:metm}\includegraphics[scale=0.24]{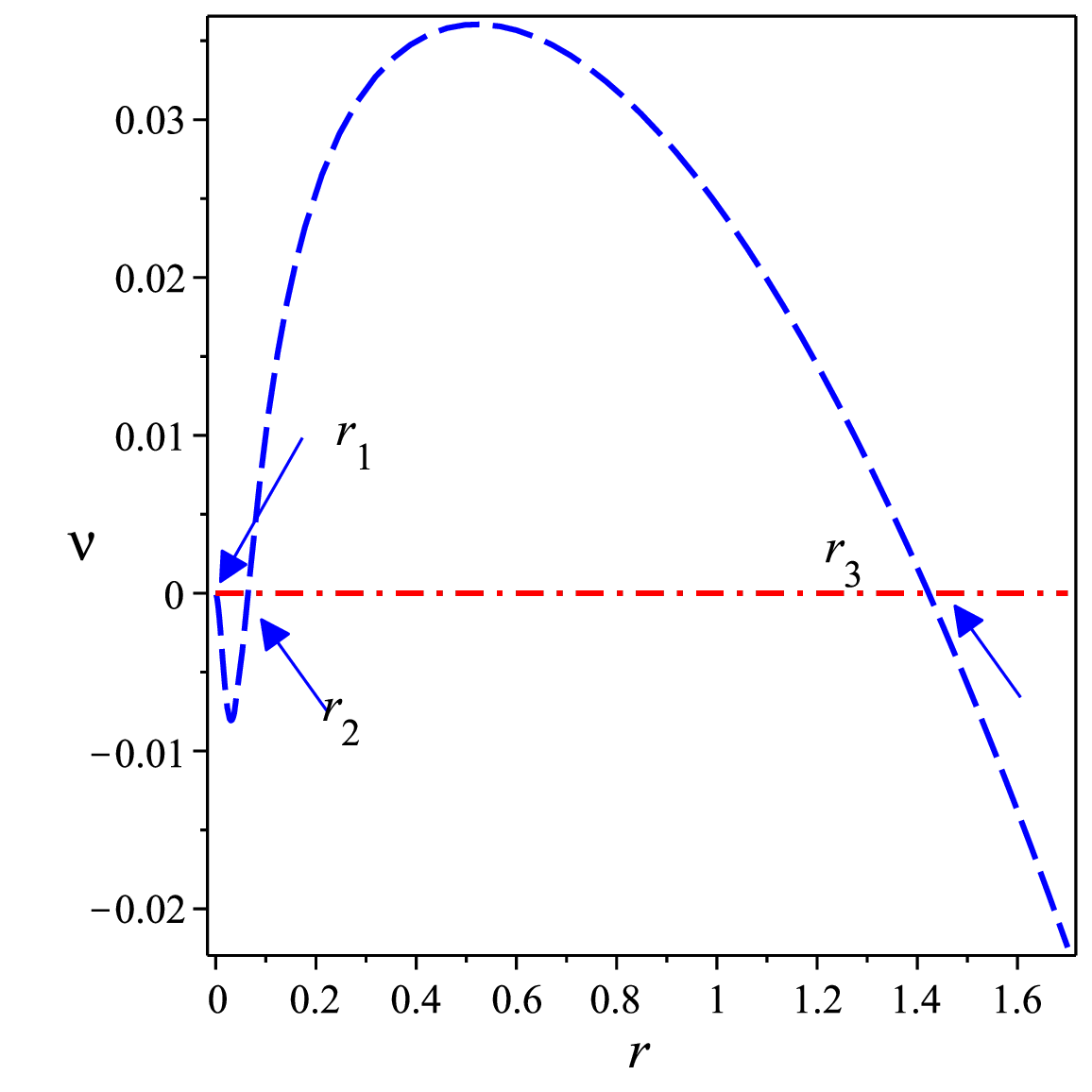}} 
\subfigure[~]{\label{fig:tempm}\includegraphics[scale=0.24]{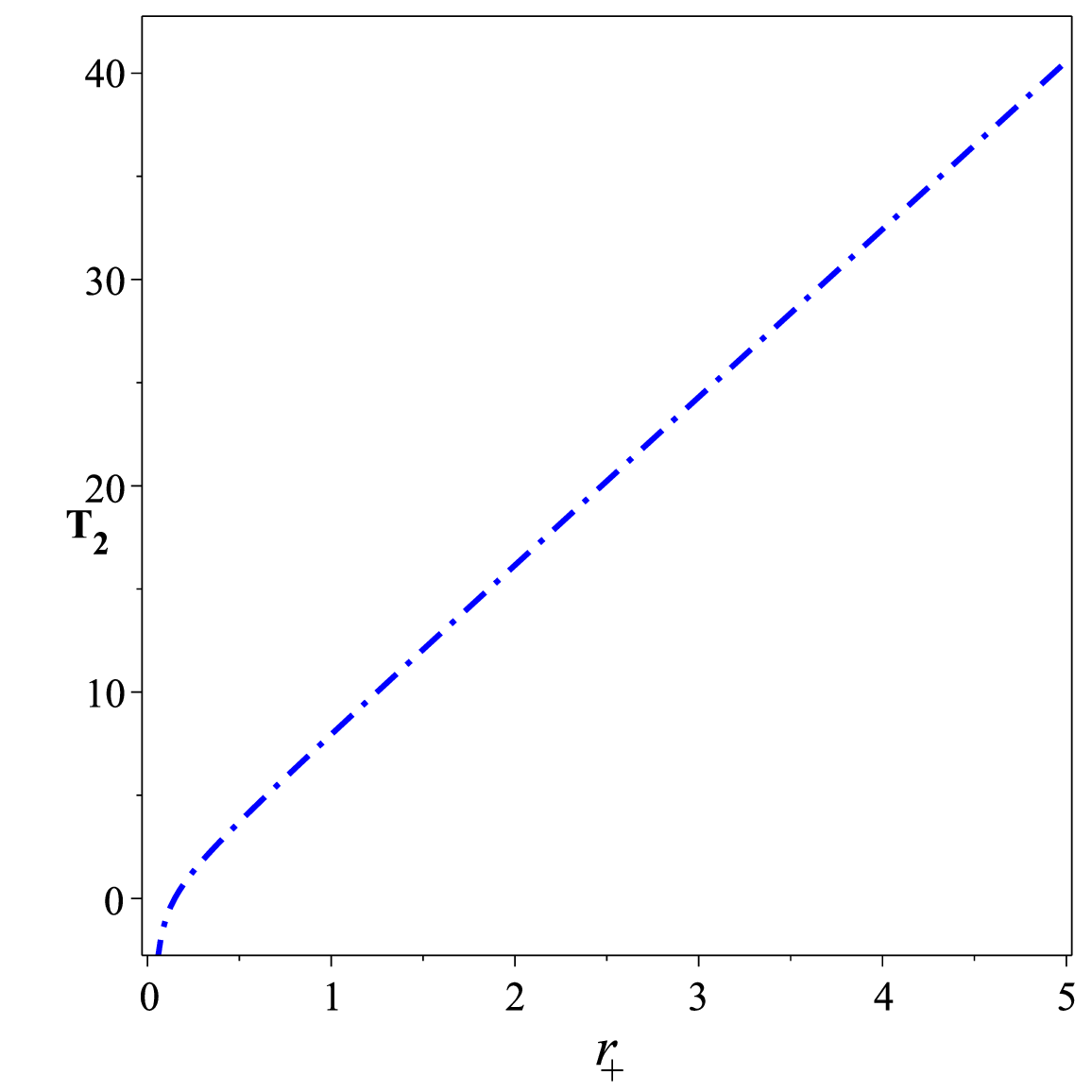}}
\subfigure[~]{\label{fig:heatm}\includegraphics[scale=0.24]{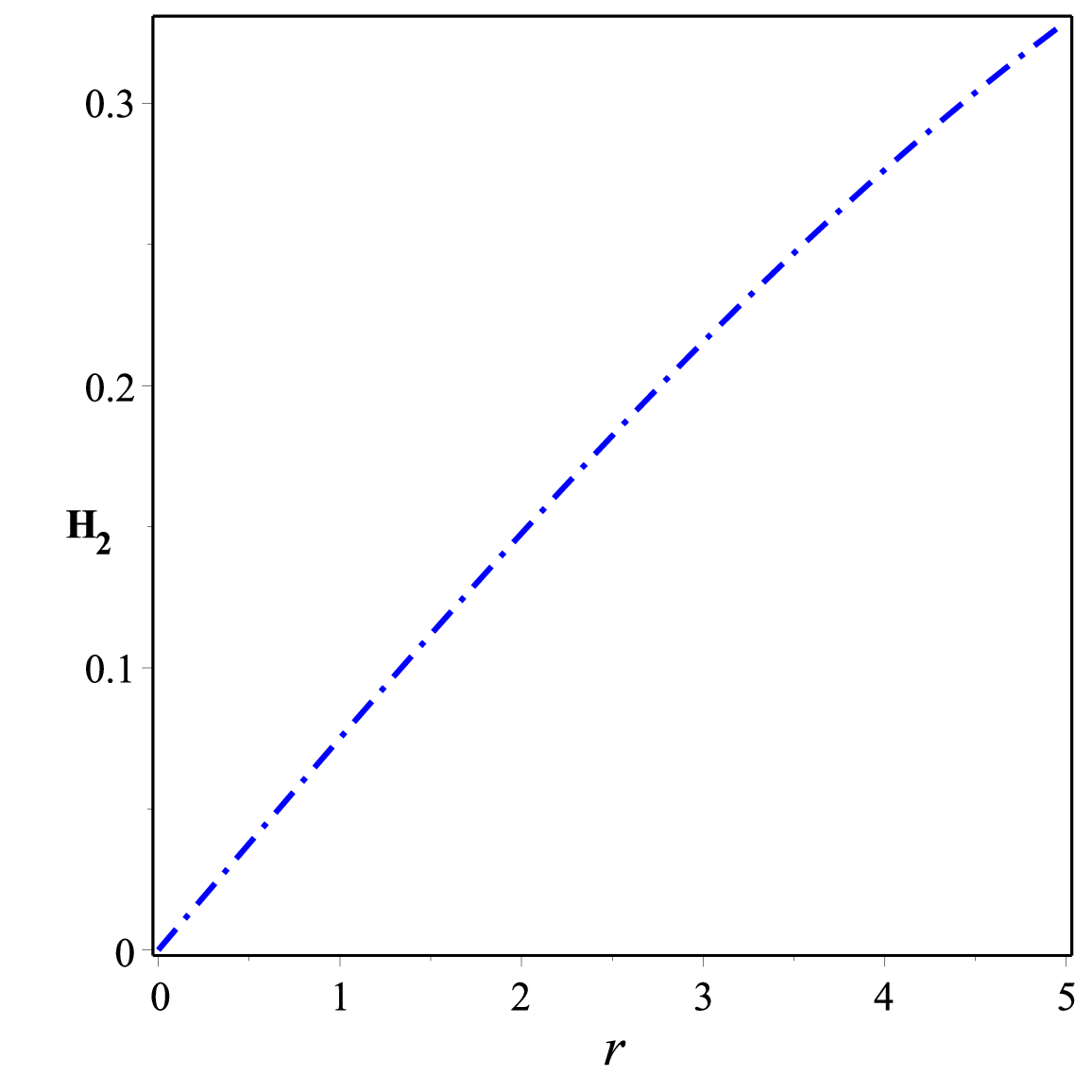}}
\caption[figtopcap]{\small{  \subref{fig:metm} The   metric     \eqref{metric}  of the multi-horizon solution.   \subref{fig:tempm} The Hawking temperature of the multi-horizon solution.   \subref{fig:heatm} The   heat capacity  of the multi-horizon solution. In all graphs we have used     $\Lambda=-0.1$, $\varphi=-0.1$, $m=0.01$.}}
\label{Fig:4}
\end{figure}

 We close our analysis by examiining the case where the charged black-hole solution exhibits multiple horizons. In order to achieve this we use the solution  \eqref{metric} where $\varphi$ takes negative values. In this case the behavior of $g_{rr}$ is shown in  Fig. \ref{Fig:4}\subref{fig:metm}. Additionally,  in order to check the viability of this case, we follow the same  procedure presented above and we     calculate   thermodynamical quantities and the effective potential. In Figs.   \ref{Fig:4}\subref{fig:tempm}  and \ref{Fig:4}\subref{fig:heatm} we present the corresponding Hawking temperature and heat capacity. As we observe,  the multi-horizon solution is stable too.

 \section{Discussion and Conclusions}
 \label{Conclusions}
 
In this study   we have constructed and analyzed novel exact charged black-hole solutions in (2+1) dimensions within the framework of  $f(Q)$ gravity. Our approach goes beyond general relativity (GR) by incorporating cubic non-metricity corrections, characterized by the function $f(Q) = -\Lambda - Q + \frac{1}{2}\alpha Q^2 + \frac{1}{3}\beta Q^3$. This formulation enables the derivation of richer geometrical and physical structures, particularly in lower-dimensional gravitational models.

We have demonstrated that the uncharged solution derived from this cubic $f(Q)$ model coincides with the well-known BTZ black hole, with the model parameters appropriately tuned to recover GR. This serves as a consistency check and confirms that the modified theory can reproduce classical solutions under specific limits.

More significantly, the derived charged solution   exhibits several novel features not attainable within GR. The solution is asymptotically Anti-de Sitter (AdS) and cannot be continuously deformed into any known GR solution, emphasizing its intrinsically  higher-order character. The metric describes a black hole with a logarithmic correction term and polynomial contributions dependending on $r^{2/3}$ and $r^{4/3}$, arising from the non-metricity-induced modifications. The resulting geometry admits configurations with two horizons, one degenerate horizon, or a naked singularity, depending on the values of the integration constants and model parameters.

We performed a detailed analysis of curvature and non-metricity invariants, showing  that the central singularity is significantly softened compared to GR, with divergence rates of the form $R, R_{\mu\nu}R^{\mu\nu} \sim r^{-2/3}$, instead of being proportional to $r^{-2}$, as in standard GR charged black holes. This indicates that higher-order non-metricity terms partially regularize the geometry       near the origin.

As a next step we derived and investigated various thermodynamic quantities, including Hawking temperature, entropy, and heat capacity. The temperature remains positive for a wide range of horizon radii, and  the entropy modified by the nontrivial $f_Q$ factor remains positive   too. The heat capacity is also positive, indicating that the black hole is thermodynamically stable. However, no critical behavior or phase transitions are observed in the P-V diagram, and the solution mimics the one-phase behavior typical of an ideal gas, suggesting the absence of Van der Waals-like thermodynamics.

Furthermore, we analyzed the geodesic structure and effective potential, revealing the existence of stable photon orbits and the influence of the cubic correction on the motion of test particles. The effective potential structure exhibits deviations from GR, confirming that the higher-order non-metricity terms significantly affect the particle dynamics near the black hole.
Lastly, we showed that within this framework multi-horizon black hole configurations are possible, by adjusting the scalar field parameter $\phi$. These solutions also exhibit thermal stability, as confirmed by positive temperature and heat capacity profiles.

In conclusion, our findings reveal the rich structure of cubic $f(Q)$ gravity in lower-dimensional spacetimes, and marks its efficiency for generating physically viable and thermodynamically stable black-hole solutions. The charged AdS black hole obtained here stands as a distinctive feature of higher-order non-metricity theories, with no GR analogue, which could further explored in  holography, quantum gravity, and gravitational thermodynamics in non-Riemannian geometries. Additionally, it would be interesting to extend the analysis in rotating cases, investigate quantum aspects such as greybody factors and quasinormal modes, or extract constraints from observational data. These studies lie beyond the scope of the present work, and will be examined in future projects.

\bibliography{JRPHSRef}
\end{document}